\documentclass[10pt]{dis03}
\usepackage{epsfig,amsmath}

\begin{document}

\title{NEW CHARM(ONIUM) RESULTS FROM CDF}

\author{ARND MEYER (for the CDF Collaboration) \\
III. Physikalisches Institut A, RWTH Aachen, Physikzentrum \\
52056 Aachen, Germany\\
E-mail: meyera@fnal.gov}

\maketitle

\begin{abstract}
\noindent
  After many upgrades to the CDF detector and to the accelerator complex,
  Run II began in April 2001. The new detector has improved capabilities for
  charm physics, and first results from the analysis of early Tevatron
  Run II data are reported here.
\end{abstract}

\section{Introduction}

  After the very successful Tevatron Run I (1992-1995), colliding protons and antiprotons
  with a center of mass energy of $\sqrt s = 1.8\:\mbox{TeV}$, the accelerator and the
  colliding beam experiments CDF and D\O\ were upgraded for Run II. On the accelerator side,
  the beam energy has been increased from $900\:\mbox{GeV}$ to $980\:\mbox{GeV}$. A new
  proton storage ring, the ``Main Injector'', has been built, and the same tunnel houses
  the ``Recycler'' storage ring, which will improve the rate at which antiprotons can be
  accumulated, and may also be utilized to reuse antiprotons after recovering them from
  the Tevatron. The bunch spacing in the Tevatron has been reduced from $3.5\:\mu\mbox{s}$
  to $396\:\mbox{ns}$. The results shown below are based on early Run II
  data corresponding to an integrated luminosity of up to $69\:\mbox{pb}^{-1}$,
  i.e.~comparable to the Run I data set ($\int {\cal L}\, dt \simeq 110\:\mbox{pb}^{-1}$).

  The CDF detector \cite{TDR} is designed for general purpose use, with a large
  tracking system inside a uniform $1.4\:\mbox{T}$ solenoidal magnetic field, $4\pi$
  calorimetry, and a muon detection system. The entire tracking system
  has been replaced for Run~II, consisting of a Silicon tracking system
  with typically seven layers at radii of $1.5\:\mbox{cm}$ to $28\:\mbox{cm}$, and
  the Central Outer Tracker (COT),
  an open cell drift chamber for the precise momentum measurement of charged tracks,
  using up to 96 space points. Between the COT and the solenoid is the new Time-of-Flight detector
  (TOF), which together with the momentum measurement
  provides particle identification by determining a particle's mass.
  The three-level trigger system and the data acquisition system \cite{daq} have been
  significantly enhanced for Run~II.
  Traditionally triggering on heavy flavor physics at hadron colliders relies on a
  lepton signature;
  examples are the decay of the $J/\psi$ into $\mu^+\mu^-$ or semileptonic $b$ decays.
  With the Silicon Vertex Tracker (SVT) \cite{SVT} at the second trigger level CDF has
  introduced a novel method to
  obtain heavy flavor decays. The SVT uses COT tracks as seeds to a parallelized
  pattern recognition in the Silicon vertex detector. The following linearized track
  fit returns track parameters with nearly offline resolution on a time scale of
  $15\:\mu\mbox{s}$. Originally designed to select hadronic
  $B$ decays the SVT also collects a large sample of charm hadrons. All of the
  measurements shown here, with the exception of the $J/\psi$ cross section, are based
  on data samples collected with the SVT.

\section{Prompt \boldmath $D$ \unboldmath Meson Cross Sections}

  There is no published measurement of the charm cross section from the Tevatron.
  With the advent of the SVT this measurement is possible in Run II, and it
  is of theoretical interest due to the larger than expected beauty
  cross sections compared to next-to-leading order QCD calculations.
  The measurement shown here, based on the analysis of $5.8\:\mbox{pb}^{-1}$ of
  data, makes use of four fully reconstructed decay modes:
  $D^0 \rightarrow K^-\pi^+$,
  $D^{\ast +} \rightarrow D^0\pi^+$ with $D^0 \rightarrow K^-\pi^+$,
  $D^+ \rightarrow K^-\pi^+\pi^+$, and
  $D_s^+ \rightarrow \phi\pi^+$ with $\phi \rightarrow K^+K^-$.

\begin{figure}[tb]
   \setlength{\unitlength}{1cm}
   \begin{picture}(12.01,7.0)(0.0,0.0)
   \put(0.1,3.3){\psfig{file=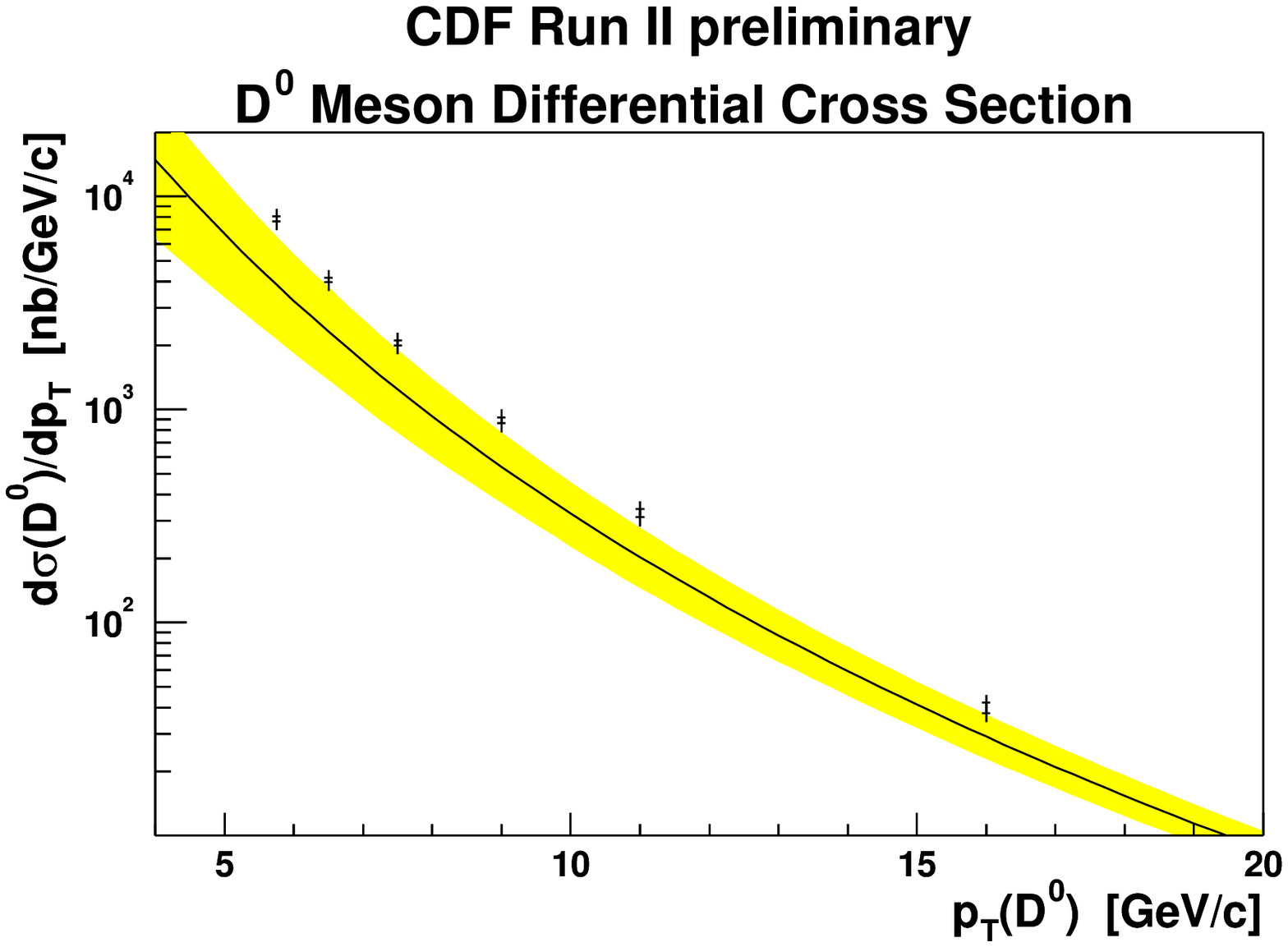,width=6.0cm}}
   \put(6.5,3.3){\psfig{file=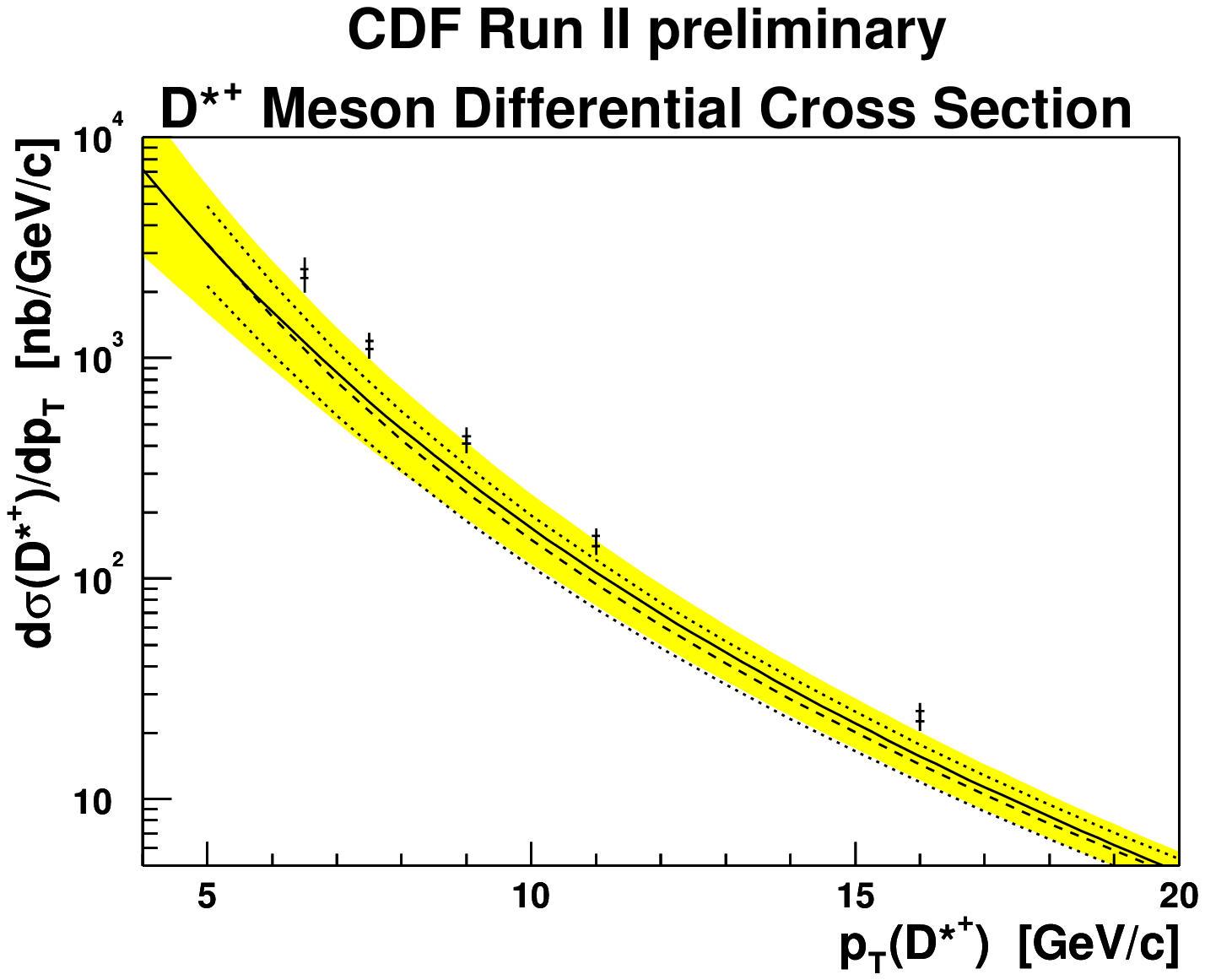,width=6.0cm,height=4.1cm}}
   \put(0.1,-0.8){\psfig{file=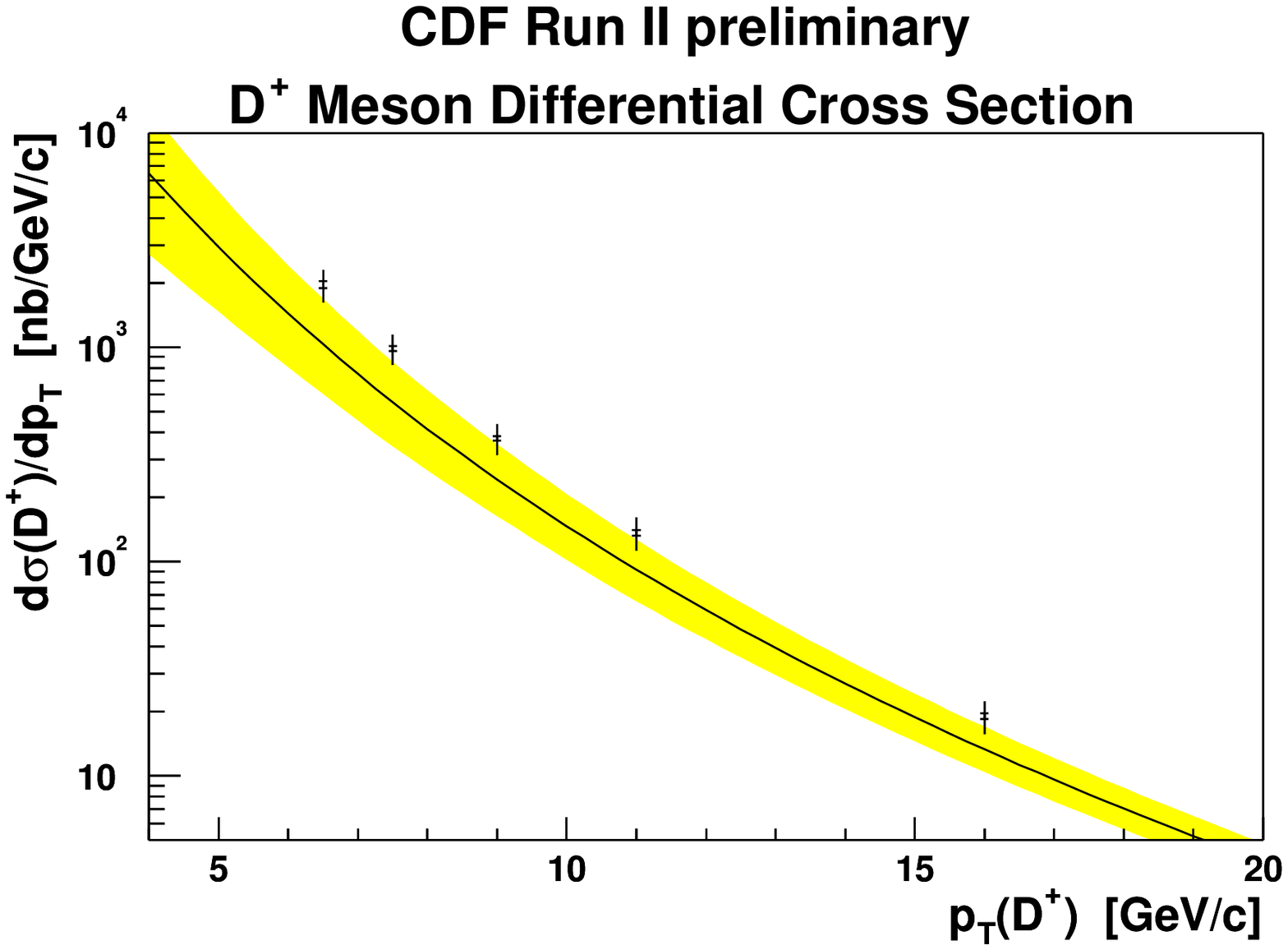,width=6.0cm}}
   \put(6.5,-0.8){\psfig{file=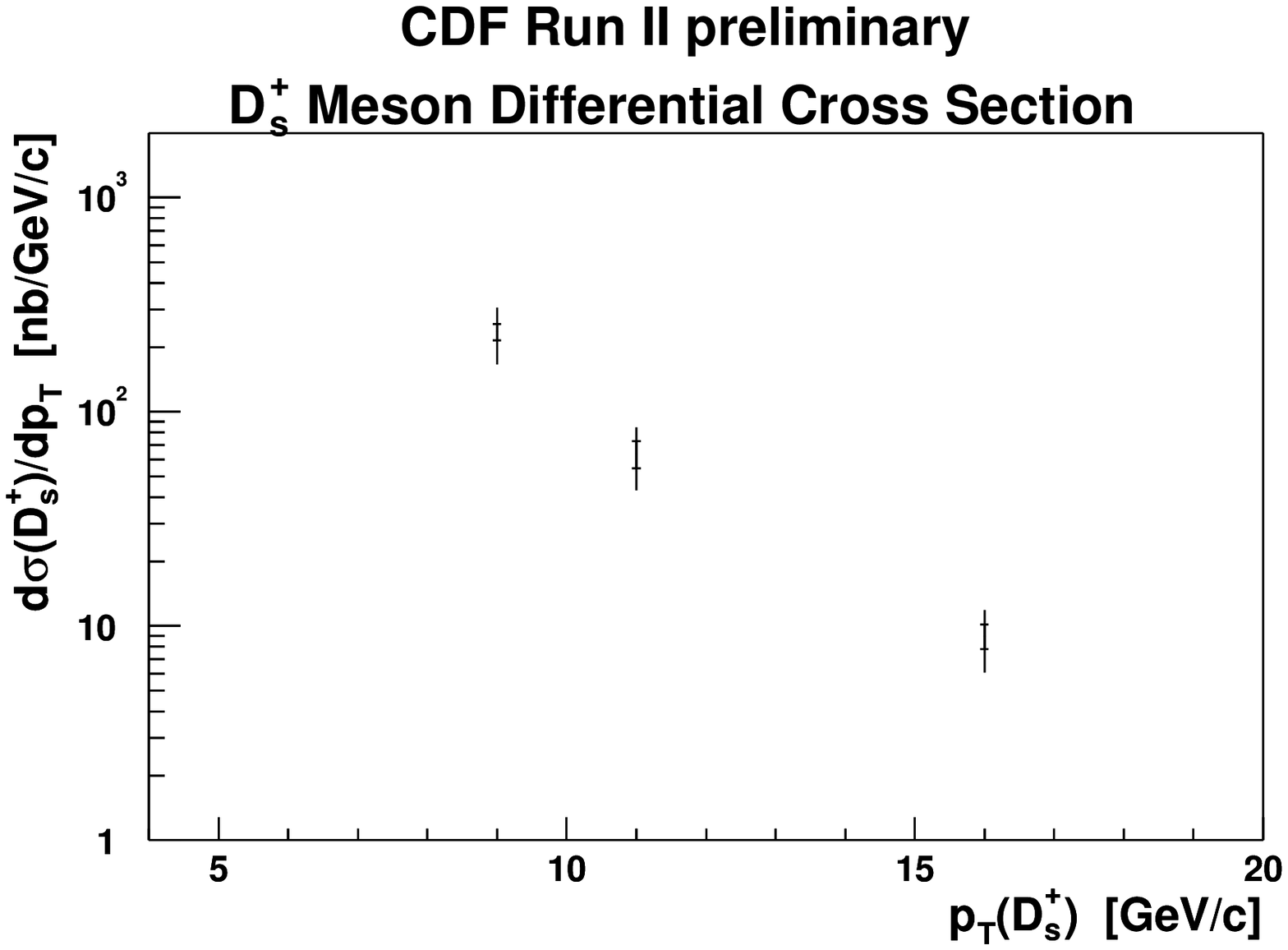,width=6.0cm}}
   \end{picture}
   \vspace*{8pt}
   \caption{Differential cross section for the four reconstructed $D$ mesons. Theoretical
            predictions from Kniehl {\em et al.} (dashed line with dotted line indicating
            uncertainty)
            and Cacciari {\em et al.} (full line with shaded band as uncertainty) are
            compared to the data. The theoretical uncertainties are based on varying the 
            renormalization and factorization scales independently by factors of 0.5 to 2.}
   \label{cxsec}
  \end{figure}

  The contributions from prompt and secondary charm are separated by utilizing
  the impact parameter distribution of the reconstructed $D$ meson samples.
  Mesons originating from $B$ decays exhibit a large impact parameter. A fit
  to the impact parameter distribution yields prompt production fractions
  of $88.6\pm 0.4 \mbox{(stat)} \pm 3.5 \mbox{(sys)}$\%, $88.1\pm 1.1\pm 3.9$\%,
     $89.1\pm 0.4\pm 2.8$\%, and $77.3\pm 3.8\pm 2.1$\% for
  $D^0$, $D^+$, $D^+$, and $D_s^+$, respectively, averaged over the full
  analyzed $p_T$ range.
  The measured prompt differential cross sections are shown in Fig.~\ref{cxsec}.
  They are compared to a next-to-leading order QCD calculation \cite{kniehl}, and a
  fixed order next-to-leading log calculation \cite{fonll}. The calculations
  are lower than, but compatible with the data.

\section{ \boldmath $m_{D_s^+}-m_{D^+}$ \unboldmath Mass Difference}

  The measurement of the $m_{D_s^+}-m_{D^+}$ mass difference provides a
  test for Heavy Quark Effective Theory and lattice QCD. While many
  precision measurements of meson masses can be expected from Run II,
  the analysis shown here could already be carried out with a modest amount of
  luminosity.



\begin{figure}[tb]
   \setlength{\unitlength}{1cm}
   \begin{picture}(12.01,3.8)(0.0,0.0)
   \put(0.3,-0.8){\psfig{file=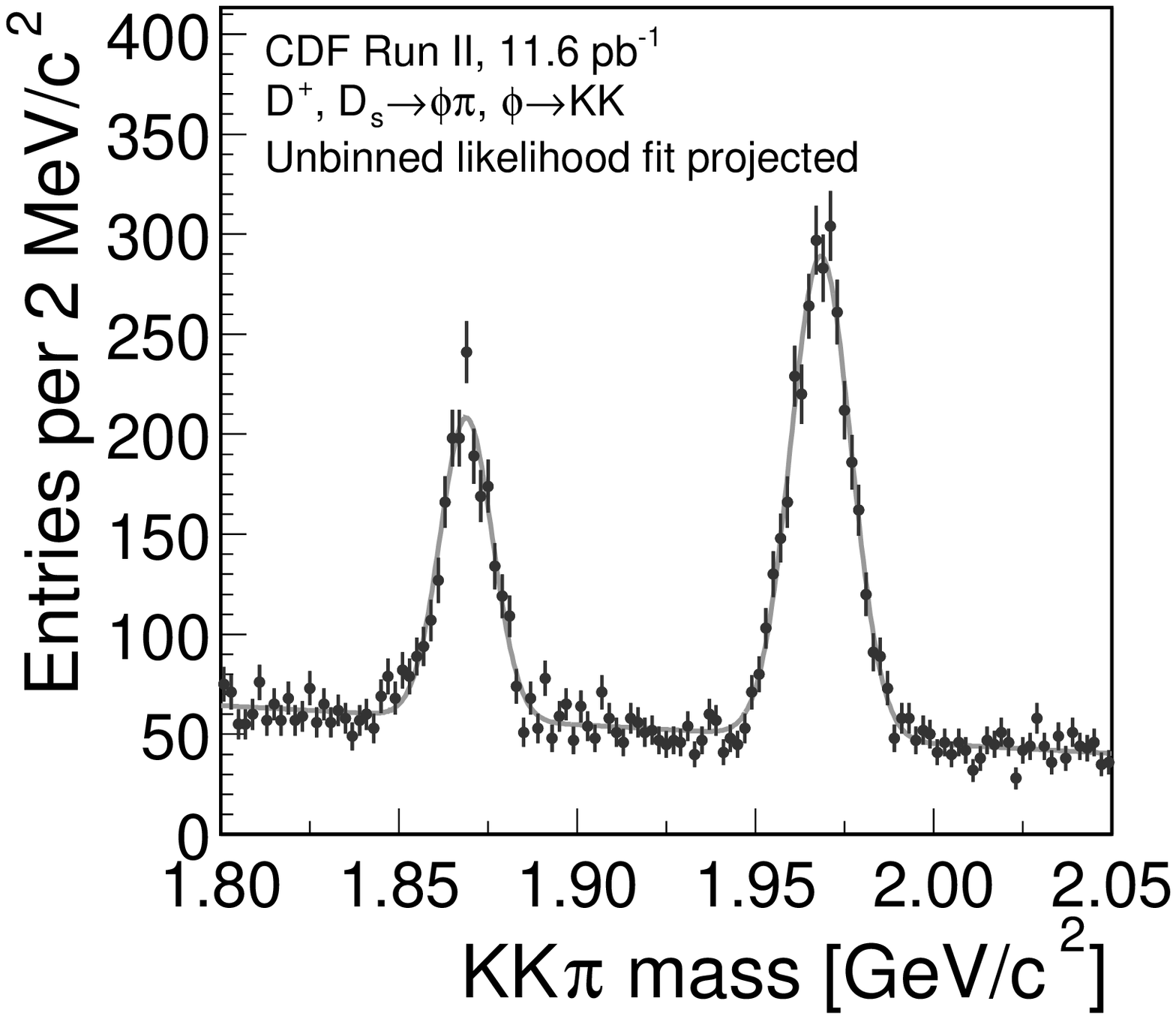,width=5.5cm}}
   \put(6.8,-0.7){\psfig{file=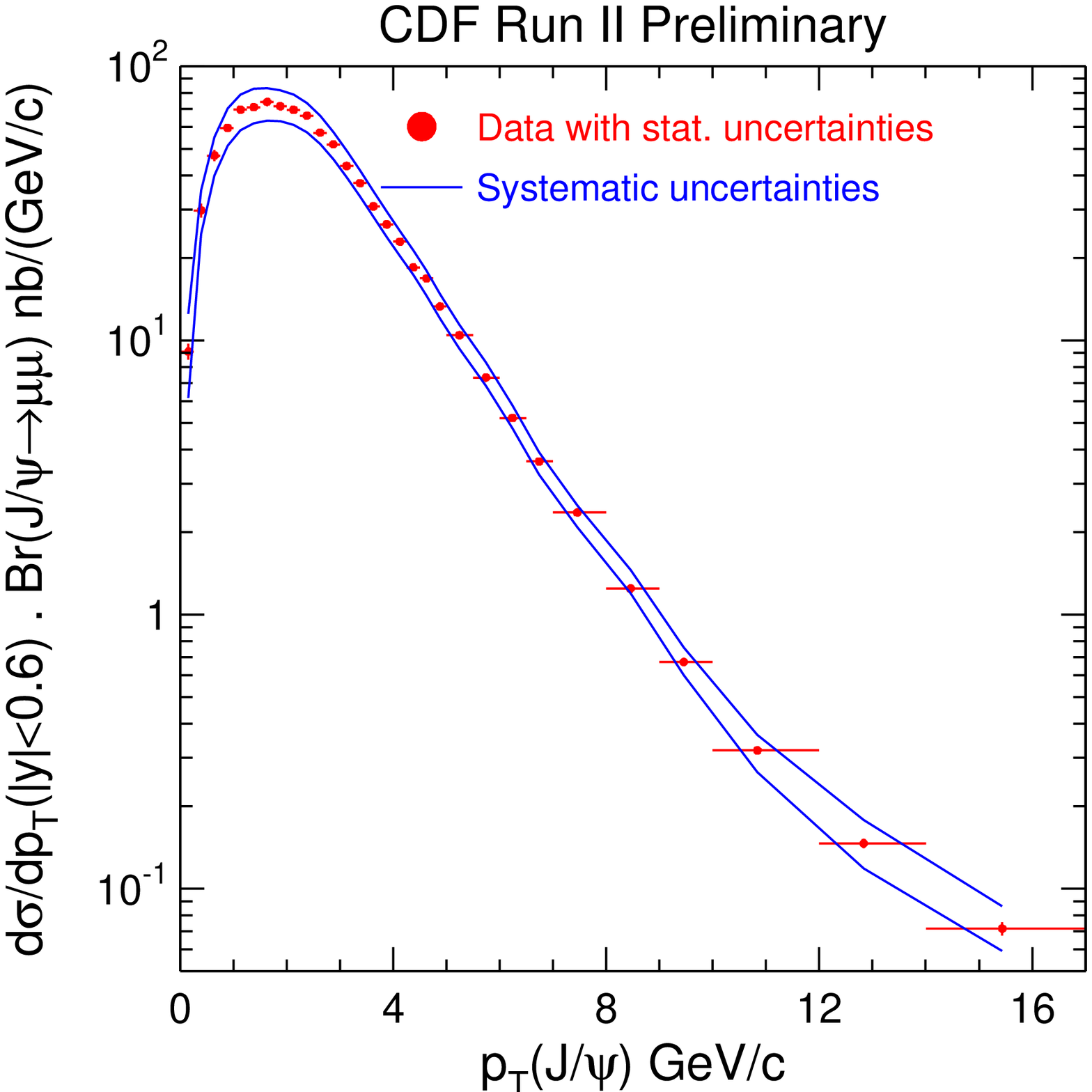,width=5.3cm}}
   \end{picture}
   \vspace*{8pt}
   \caption{{\em Left:} The reconstructed $D^+ \rightarrow \phi\pi^+$ and $D_s^+ \rightarrow \phi\pi^+$ mass
            distribution. {\em Right:} The inclusive $J/\psi$ cross section
            differential in $p_T$ for $|y(J/\psi)|<0.6$.}
   \label{dsignals}
  \end{figure}

  The mass difference measurement relies on $D_s^+$ and $D^+$ decays into
  $\phi\pi^+$ with $\phi\rightarrow K^+K^-$, as shown in Fig.~\ref{dsignals}.
  Using the same decay mode has the advantage of cancelling systematic uncertainties.
  The unbinned likelihood fit of the mass spectrum results in
  $m_{D_s^+}-m_{D^+} = 99.41 \pm 0.38 \mbox{(stat)} \pm 0.21 \mbox{(sys)}\:\mbox{MeV}$,
  with uncertainties comparable to the world average \cite{pdg}. This measurement
  constitutes the first published Run II result from CDF \cite{mass}.

\section{Cabibbo Suppressed Decays and CP Violation}

\begin{figure}[tb]
   \setlength{\unitlength}{1cm}
   \begin{picture}(12.01,2.8)(0.0,0.0)
   \put(-0.2,-0.8){\psfig{file=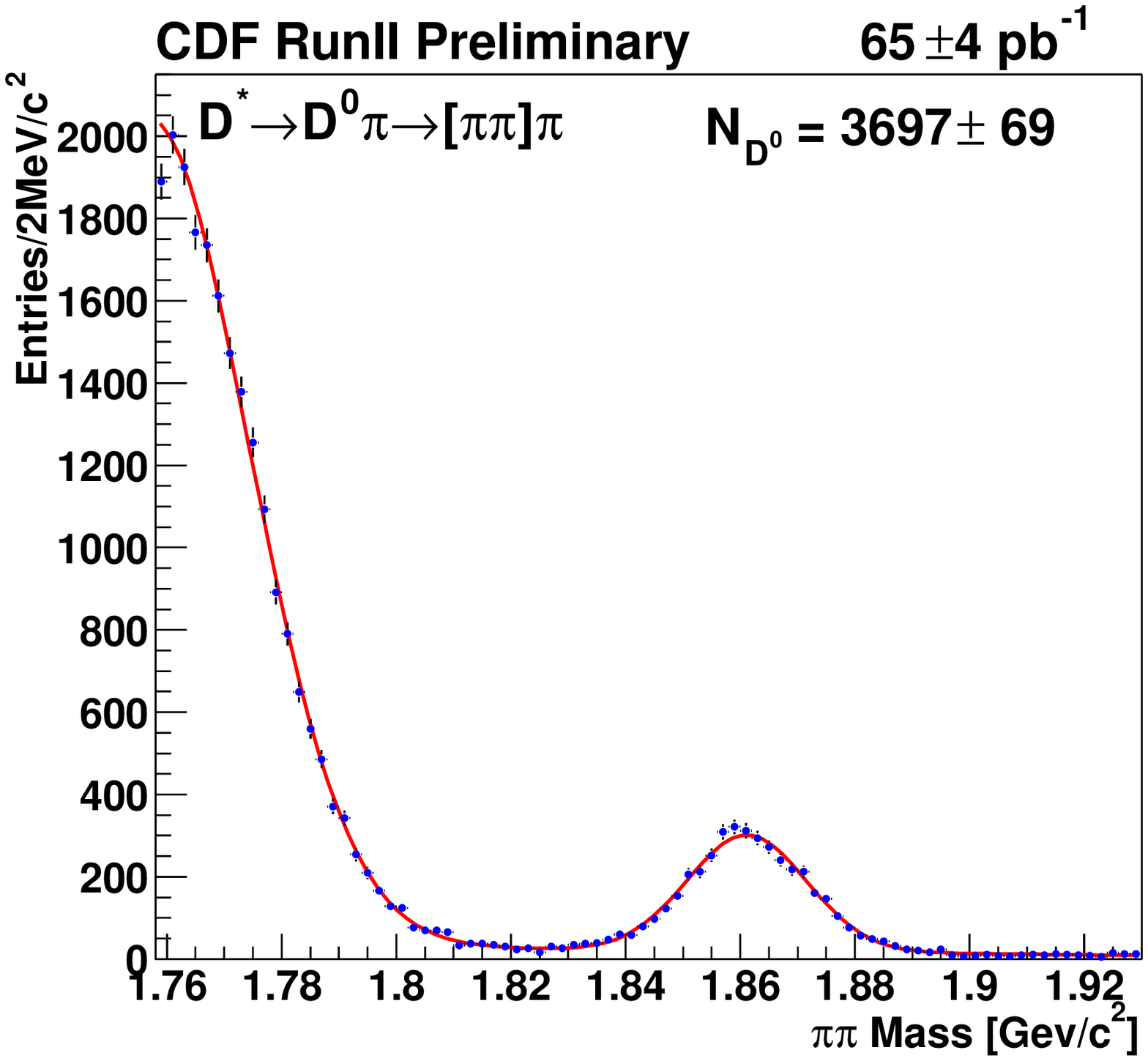,width=4.5cm}}
   \put(4.0,-0.8){\psfig{file=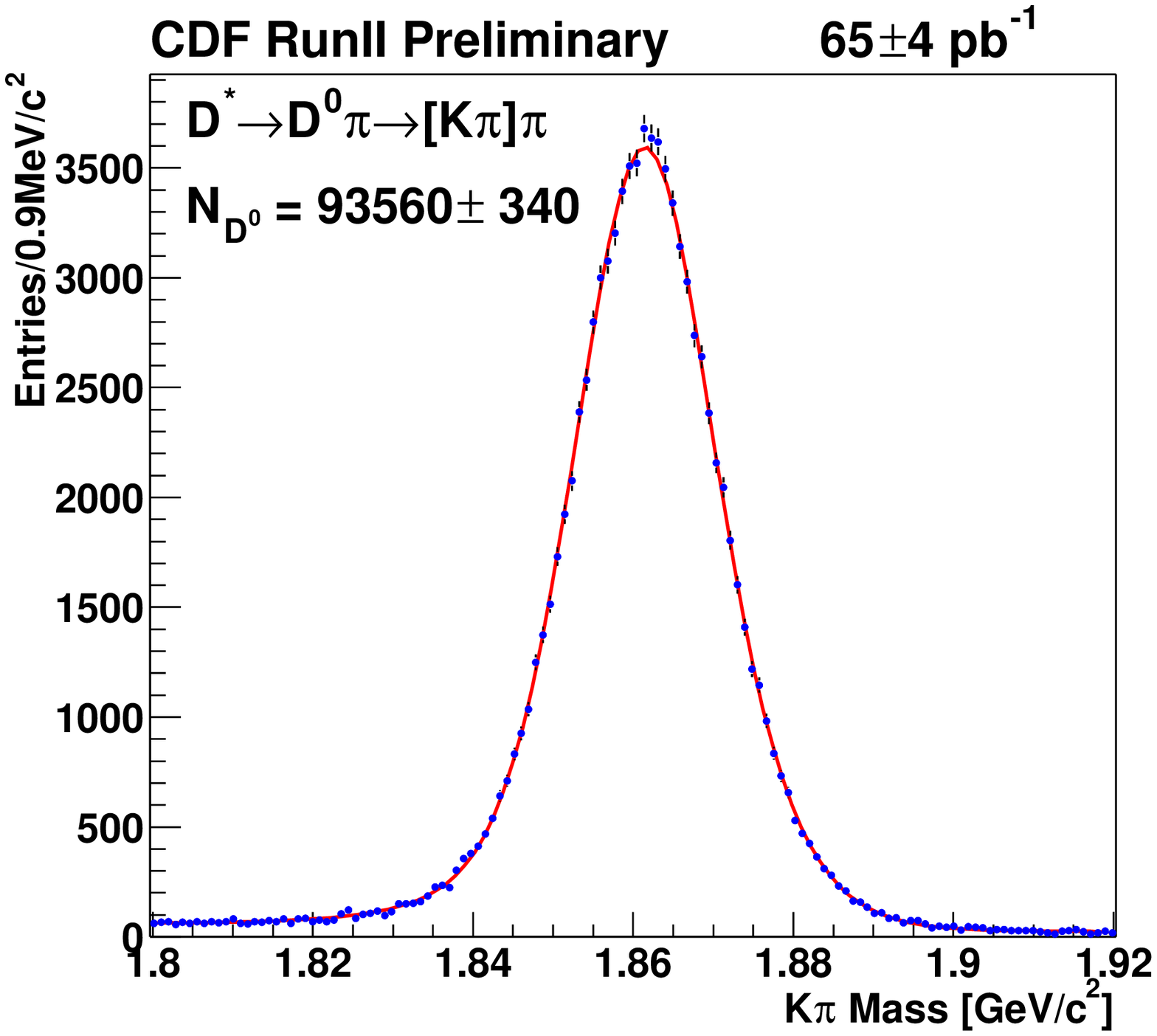,width=4.5cm}}
   \put(8.2,-0.8){\psfig{file=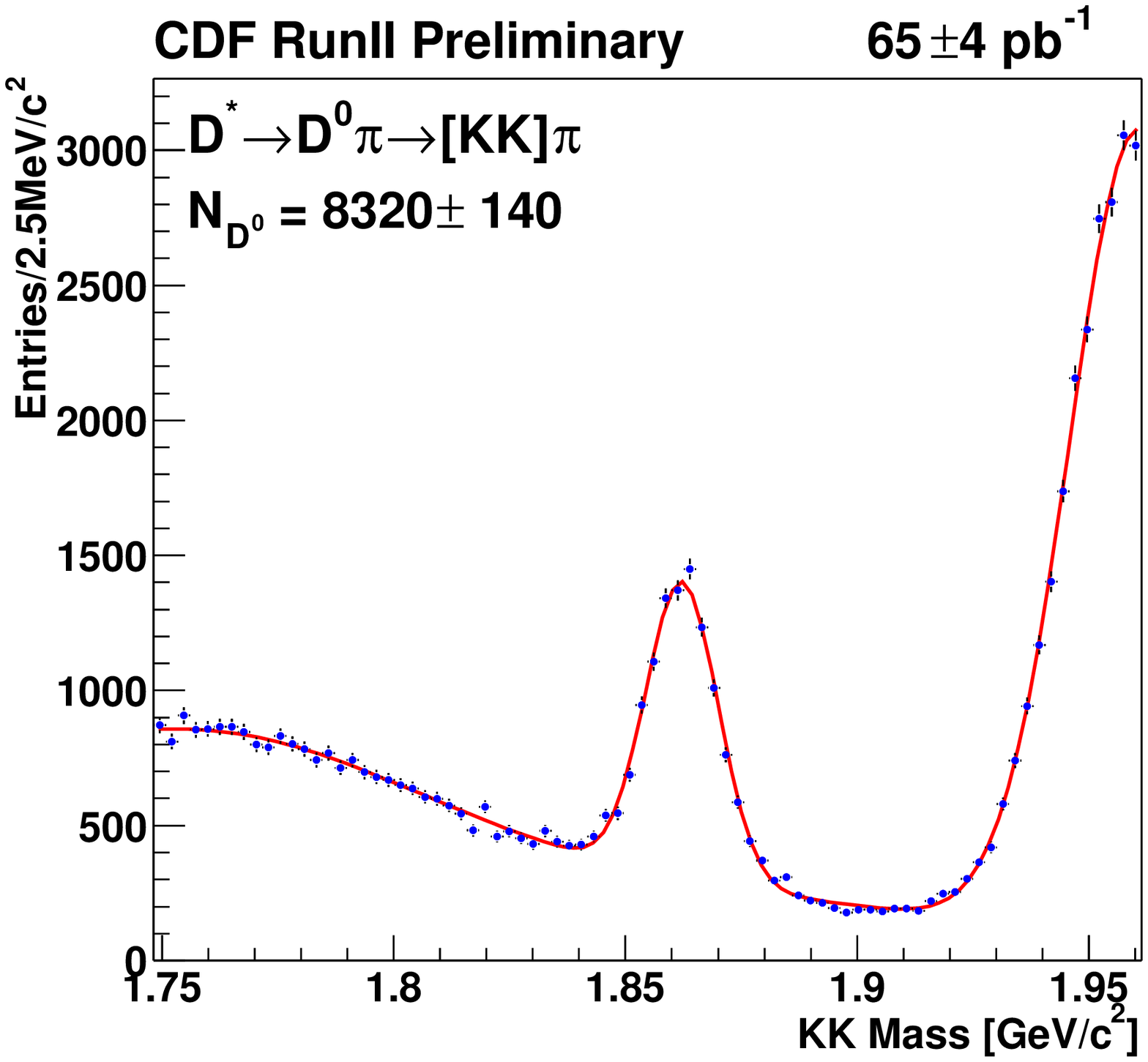,width=4.5cm}}
   \end{picture}
   \vspace*{8pt}
   \caption{Reconstructed $D^0$ decay modes for branching ratio and CP asymmetry
            measurements. Note that in the $KK$ (right) and $\pi\pi$ (left) modes
            the reflections are well separated.}
   \label{cabibbo}
  \end{figure}

  Utilizing the large sample of $D^0$ mesons in $65\:\mbox{pb}^{-1}$ integrated
  luminosity collected with the secondary vertex trigger, relative branching
  fractions are measured,
  $\frac{\Gamma(D^0\rightarrow K^+K^-)}{\Gamma(D^0\rightarrow K^+\pi^-)} =
         9.38 \pm 0.18 \mbox{(stat)}\pm 0.10 \mbox{(sys)}$\%
  and
  $ \frac{\Gamma(D^0\rightarrow \pi^+\pi^-)}{\Gamma(D^0\rightarrow K^+\pi^-)} =
        3.686 \pm 0.076 \mbox{(stat)}\pm 0.036 \mbox{(sys)}$\%,
  comparing favorably with the current best measurement \cite{focus}. In the
  analysis, the $D^0$ candidate is combined with a charged slow pion
  to form a $D^\ast$ meson; in this way, backgrounds are reduced, and the charge
  of the slow pion from the $D^\ast$ decay serves as an unbiased tag of the
  $D^0$ flavor. Examples of the reconstructed decays are shown in Fig.~\ref{cabibbo}.

  The CP violating decay rate asymmetries
  $A=\frac{\Gamma(D^0\rightarrow f) - \Gamma(\bar{D^0}\rightarrow f)}{
           \Gamma(D^0\rightarrow f) + \Gamma(\bar{D^0}\rightarrow f)}$
  are also measured. It is found that
     $A(D^0\rightarrow K^+K^-) = 2.0 \pm 1.7 \mbox{(stat)}\pm 0.6\mbox{(sys)}$\% and
     $A(D^0\rightarrow \pi^+\pi^-) = 3.0 \pm 1.9 \mbox{(stat)}\pm 0.6\mbox{(sys)}$\%,
  comparable to previous measurements \cite{cleo}.

\section{Search for the FCNC Decay \boldmath $D^0 \rightarrow \mu^+\mu^-$\unboldmath }

\begin{figure}[tb]
   \setlength{\unitlength}{1cm}
   \begin{picture}(12.01,3.8)(0.0,0.0)
   \put(0.7,-0.7){\psfig{file=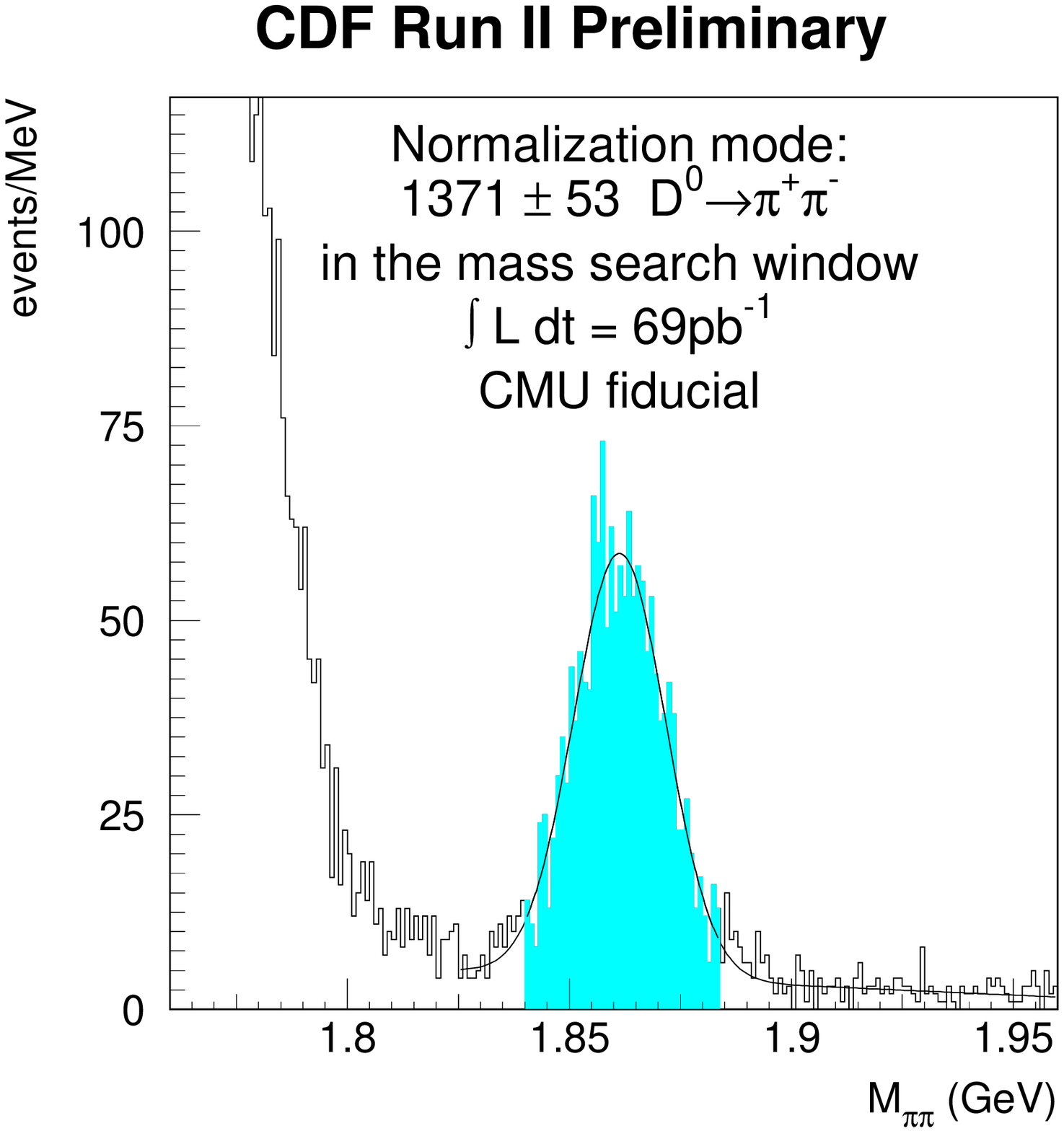,width=5.1cm}}
   \put(6.5,-0.7){\psfig{file=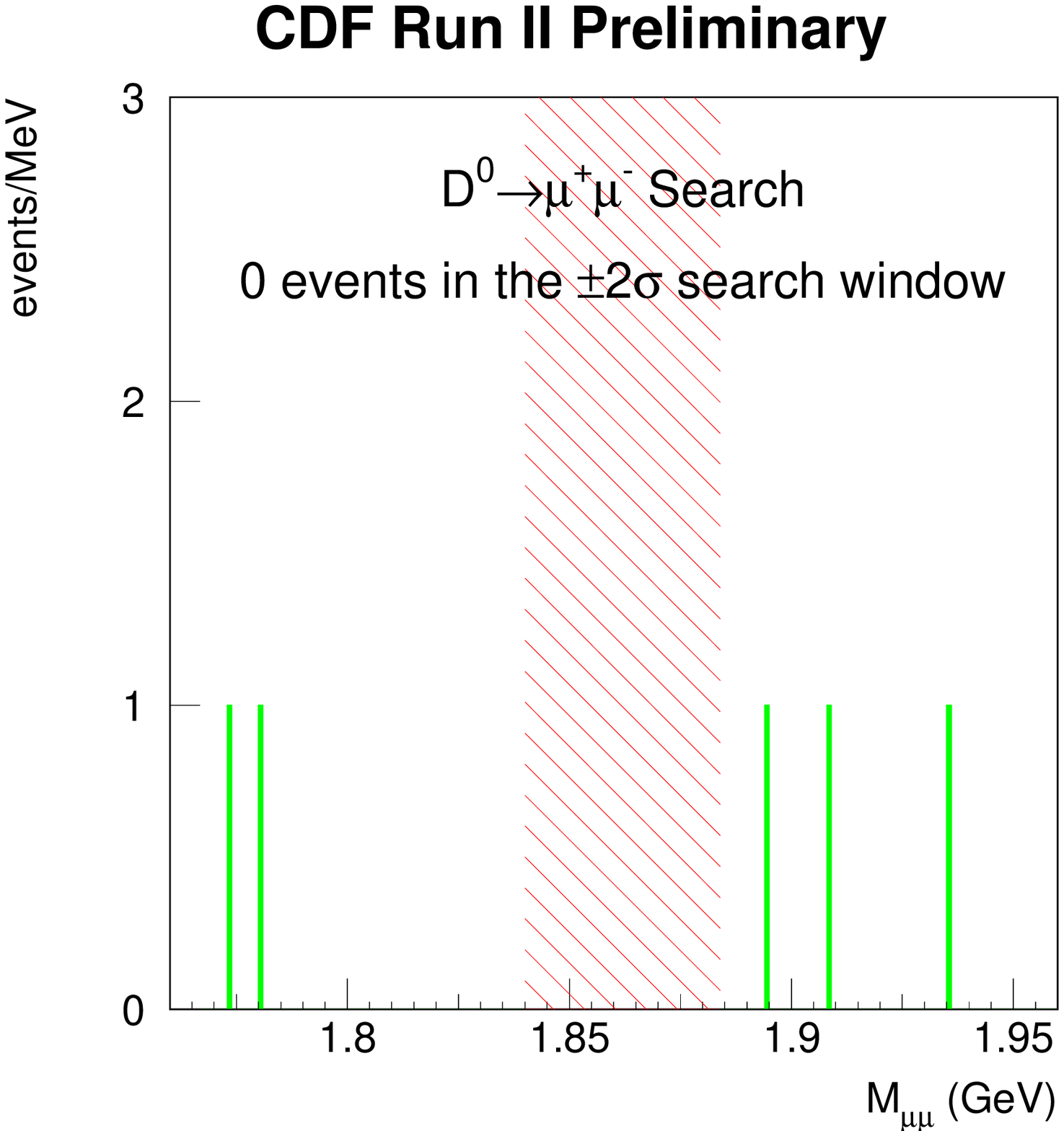,width=5.1cm}}
   \end{picture}
   \vspace*{8pt}
   \caption{Search for $D^0 \rightarrow \mu^+\mu^-$. {\em Left:} Mass spectrum for the
            normalization mode $D^0 \rightarrow \pi^+\pi^-$; the search window is
            indicated by the shaded area. {\em Right:} Dimuon candidate events; the
            search window is indicated by the hatched area.}
   \label{dmumu}
  \end{figure}

  The search for the flavor changing neutral current (FCNC) decay
  $D^0 \rightarrow \mu^+\mu^-$ is another example of an analysis that greatly
  benefits from the SVT trigger, by providing the well measured normalization
  mode $D^0 \rightarrow \pi^+\pi^-$.
  The branching ratio is ${\cal O}(10^{-13})$ in the Standard Model, but can
  be enhanced up to ${\cal B}(D^0 \rightarrow \mu^+\mu^-) \simeq 3.5\cdot 10^{-6}$
  in R-parity violating models of Supersymmetry.
  In a data sample corresponding to an integrated luminosity of $69\:\mbox{pb}^{-1}$
  no candidate event is observed (Fig.~\ref{dmumu}) with $1.7\pm 0.7$ background
  events expected. After correcting for relative acceptance an upper limit of
  ${\cal B}(D^0 \rightarrow \mu^+\mu^-) \leq 2.4\cdot 10^{-6}$ at 90\% CL
  is set, improving the current best limit \cite{mumulimit} of $4.1\cdot 10^{-6}$.

\section{Measurement of the \boldmath $J/\psi$ \boldmath Cross Section}

  One of the surprises of Run I was the direct production
  cross section for $J/\psi$ and $\psi(2S)$ mesons \cite{psirun1}, which turned out to be
  orders of magnitude larger than the theoretical expectation in the Color
  Singlet Model. While later calculations within the framework of non-relativistic
  QCD, including intermediate color octet states, are in broad agreement with the
  data, there is continued interest in the subject; in particular measurements
  of the $J/\psi$ and $\psi(2S)$ polarization appear to be in conflict with the
  theory, albeit not with convincing statistical significance.

  For Run II the muon trigger momentum thresholds at CDF were lowered to
  $\geq 1.4\:\mbox{GeV}$, thus allowing to trigger on $J/\psi$'s at rest for
  the first time. Using a data sample of $39.7\:\mbox{pb}^{-1}$ the inclusive differential
  cross section in bins of the $J/\psi$ transverse momentum has been measured for $J/\psi$ rapidities
  $|y(J/\psi)|<0.6$ (Fig.~\ref{dsignals}). The region $p_T < 5\:\mbox{GeV}$ is covered for the first
  time. The total cross section has been determined to be
  $\sigma(p\bar{p} \rightarrow J/\psi X, |y(J/\psi)|<0.6) \cdot {\cal B}(J/\psi \rightarrow \mu^+\mu^-)
   = 240 \pm 1 \mbox{(stat)} ^{+35}_{-28} \mbox{(sys)} \:\mbox{nb}$.
  In the future, the determination of the prompt production
  cross section, more precise data at large $p_T$, polarization measurements, and
  similar measurements of $\psi(2S)$, $\Upsilon(1S)$, $\Upsilon(2S)$, and
  $\Upsilon(3S)$ production will shed further light on the production mechanisms
  for heavy vector mesons.

\end{document}